\documentclass[12pt]{iopart}

\usepackage{iopams}
\usepackage{graphicx}  
\begin{document}

\title[Casimir forces between dielectric spheres]{Casimir forces between spheres
\footnote{Talk given at the CEWQO 2009 (May), Turku, Finland.}}
\author{James Babington}

\address{
Quantum Optics and Laser Science, Blackett Laboratory, \\
Imperial College London, Prince Consort Road, \\
London SW7 2AZ, U. K.}
\ead{j.babington@imperial.ac.uk}

\begin{abstract}
We discuss the calculation of Casimir forces between a collection of $N$-dielectric spheres. This is done by evaluating directly the force on a sphere constructed from a stress tensor, rather than an interaction energy. Two and three body forces between the spheres are evaluated for setups of two and three sphere systems respectively. An approximate large-$N$ limit is also obtained for the functional dependence on the number of spheres.

\end{abstract}


\section{Introduction}

In a recent set of papers~\cite{emig-2008-41,kenneth:014103,bulgac:025007,rahi-2009}, Casimir interaction energies have been evaluated for collections of compact objects interacting with different force carrying fields (electromagnetic and scalar). The approach taken has been to evaluate a suitable energy functional integral using a T-matrix, whereby an interaction energy can be deduced, normalised with respect to their energy when separated at infinity. This is a clean and appealing way of evaluating Casimir energies once the effective action and integration measures are known. See also~\cite{PhysRevE.55.4207,Barton1} for related earlier work on spherical geometry issues.

Looking more generally, one may try to evaluate Casimir interactions (forces or energies) with definitions that are not necessarily equivalent. For example, the Minkowski stress tensor is \emph{not} compatible with the Lorentz force law; it is rather the standard vacuum expression that is. The fact that the definitions lead to different forces means they are amenable to experimental testing; three body forces can be measured between dielectric spheres~\cite{PhysRevLett.92.078301} and thus discernible outcomes can be tested for the best candidate theory. Typically the differences will show up in both the scale of the forces and the higher order curvature corrections.

In this talk I summarise recent work we have done on calculating the Casimir force on a single sphere, in an $N$-sphere setup~\cite{James2009}. By using a multiple scattering approach to evaluate essentially the  classical Green's function of the configuration, we are able to evaluate the force \emph{directly} on the sphere. It is given in terms of Mie scattering coefficients together with translation matrices that map TE and TM modes between different scattering centres. This is calculated explicitly for the case of two and three sphere setups (with overall evaluation being performed in Maple). For the case where we have a large number of spheres that interact weakly, we can form a new coupling constant for the perturbative expansions and look at the dependence on $N$ rather than the details of the configurational setup.

\section{The $N$-Sphere configuration}

\begin{figure}[htbp]
\begin{center}
\includegraphics[height=7cm]{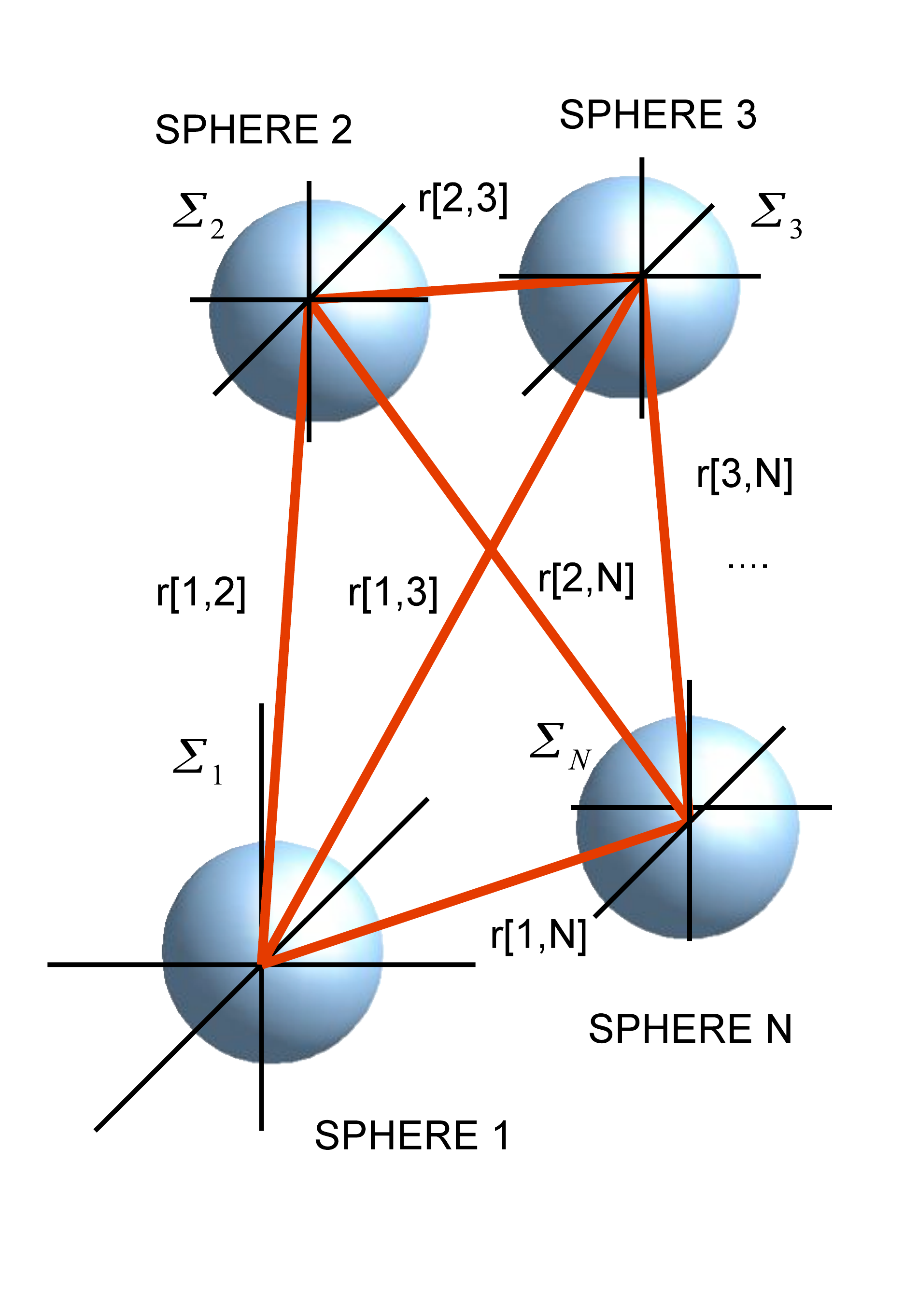}
\caption{{\bf The $N$-Sphere system consists of $N$ dielectric spheres of radii $R[1],\cdots, R[N]$ each centred on $N$ separate coordinate systems $\Sigma_1, \cdots ,\Sigma_N$, all contained in a background dielectric.}}\label{fig:NSPHERES}
\end{center}
\end{figure}

The problem we are addressing is how to calculate the force on a particular sphere as a result of all the interactions with the remaining spheres in a particular static configuration.  Pictorially, the setup we have for the configuration of $N$-spheres is shown in Figure~\ref{fig:NSPHERES}. We shall take sphere-1 to be located at the coordinate origin.

The Casimir force on sphere-1 (in the $j$-direction) due to the effects of the $N$-sphere system of differing material properties is given by
\begin{equation}
\label{eq:stressintegral}
F^j(1|N-1)=\int_{B^2} d^3x \nabla_i T^{ij}(x).
\end{equation}
The stress tensor is given by the standard vacuum expression (which is consistent with the Lorentz force law~\cite{raabe:013814})
\begin{equation}
T_{ij}(x)=\mathbf{E}_i(x)\mathbf{E}_j(x)+\mathbf{B}_i(x)\mathbf{B}_j(x)-\frac{1}{2}\delta_{ij}(|\mathbf{E}(x)|^2+|\mathbf{B}(x)|^2),
\end{equation}
where  $x \in B^2$  and it is understood that we are taking the limit for the initial and final points. We then need to evaluate the scattering correlation functions (whilst dropping the direct modes of propagation), viz
\begin{equation}
\lim_{y \rightarrow x}\mathbf{E}_i(x)\mathbf{E}_j(y)=\int_{0}^{\infty}\int_{0}^{\infty}d\omega d\omega^{\prime}\langle \mathbf{E}^{out}_{i}(x;\omega)^{\dagger}\mathbf{E}^{in}_{j}(y;\omega^{\prime})\rangle,
\end{equation}
and similarly for magnetic fields. To construct the scattering two point function we write the fields in a mode decomposition~\cite{Mackowski06081991} of spherical vector wave functions that are centred on each sphere centre. Then by applying the standard continuity equations at each of the spheres surfaces, one can calculate the out modes in terms of the in modes and scattering (Mie) coefficients. Care must be taken when evaluating effectively the classical scattering Green's tensor for the $N$ spheres. It is  necessary to choose the centre of the  sphere where the scattering takes place last as the coordinate origin to evaluate the derivative, whilst translating this to the centre of the sphere where the force is being evaluated (so as to be able to evaluate the eigenfunctions on the sphere). Assuming that the background in which we are evaluating this is filled with quantum noise such that the noise-current two point function is non-zero we find for the $N$-body force on sphere-1 (suppressing the $SO(3)$ indices)
\begin{eqnarray}
\label{eq:NSPHEREFORCE2}
\mathbf{F}[1|N-1]&=&-(-1)^N\frac{\hbar }{4\pi } R[1]
 \Im \int_{0}^{\infty}d\omega k \coth (\hbar \omega /K_BT)\langle\mathbf{1}| [ \alpha^{1}(\omega R[1]) \nonumber \\ 
& &\times 
\sum_{i=2}^{N}A^{1,i}(\mathbf{r}[1,i]) \cdot \alpha^{i}( \omega R[i] )\cdots 
\cdots \sum_{j=2}^{N}A^{i,j}(\mathbf{r}[i,j]) \cdot \alpha^{j} \nonumber \\
&&\times \sum^N_{j=2}\nabla_{\mathbf{r}[j,1]} A^{j,1}(\mathbf{r}[j,1])] j(kR[1])
h^{+}(kR[1])W(\omega R[1]) |\mathbf{1}\rangle. 
\end{eqnarray}
Here,  $\alpha^i(\omega R[i])$ are the Mie scattering coefficients in the $SO(3)$ basis for sphere $i$ with radius $R[i]$. In addition $A^{i,j}(\mathbf{r}[i,j])$ are the translation matrices mapping the modes between spheres $i$ and $j$, and the vectors $| \mathbf{1}\rangle$ give the truncation in the $L$ angular momentum quantum number (leading to a multipole type expansion). Note the explicit form of the translation matrices involve exponentials of the inter-sphere separations~\cite{James2009} and thus it is the \emph{total path length} that plays the key role in understanding the variables of the system.

\section{Two and Three Sphere Forces}

\begin{figure}[htbp]
\begin{center}
\includegraphics[height=5.5cm]{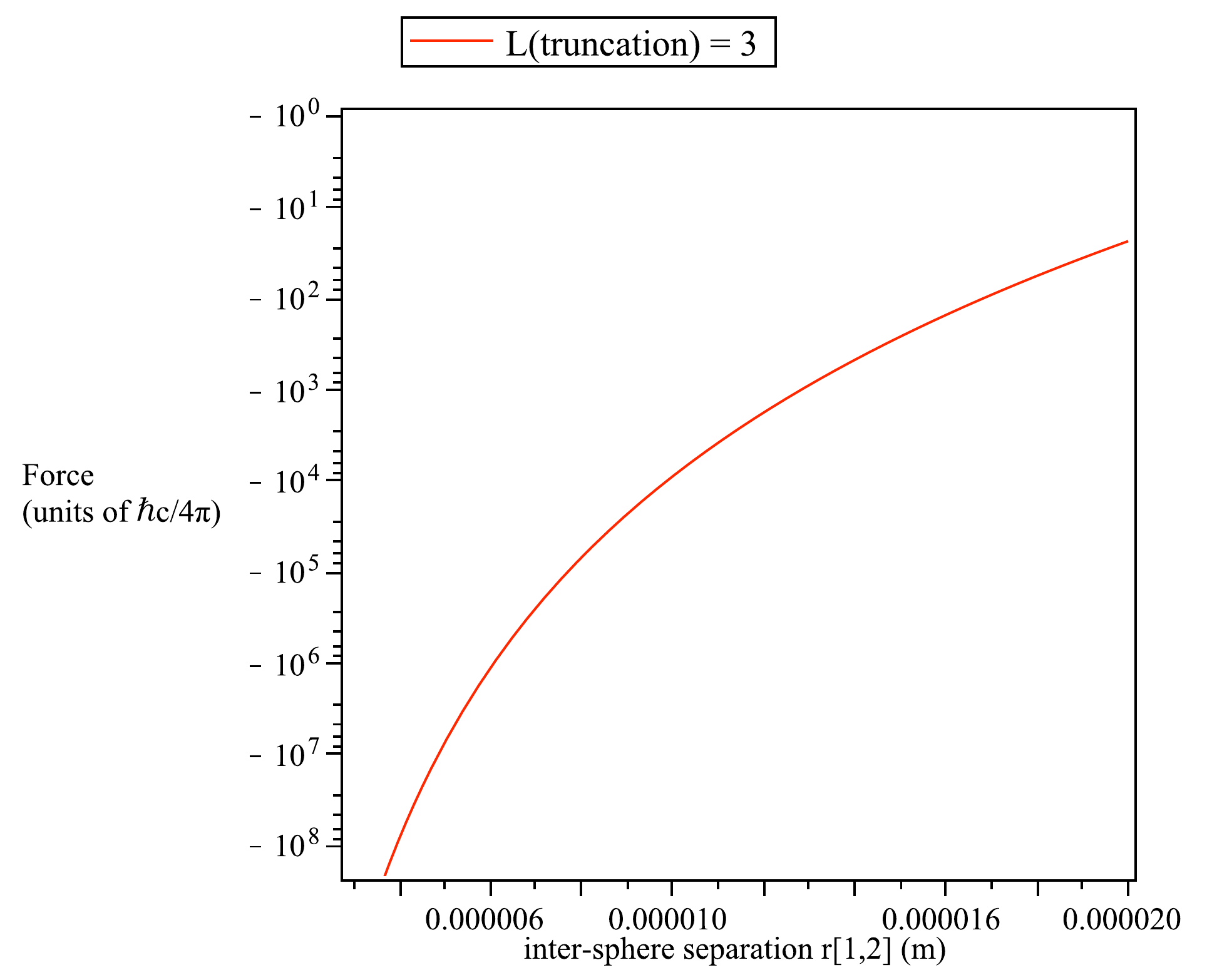}
\caption{A plot of the inter-sphere retarded force between two dielectric spheres in the empty vacuum at $T=0^{\circ} K$. Here, the relative dielectric permittivity of the polystyrene spheres is $\epsilon_1=\epsilon_2=2.6$ . The angular momentum series representation has been truncated at $L_{truncate}=3$.}\label{fig:TWOSPHERES}
\end{center}
\end{figure}

\begin{figure}[htbp]
\centering
\begin{center}
\includegraphics[height=5.5cm]{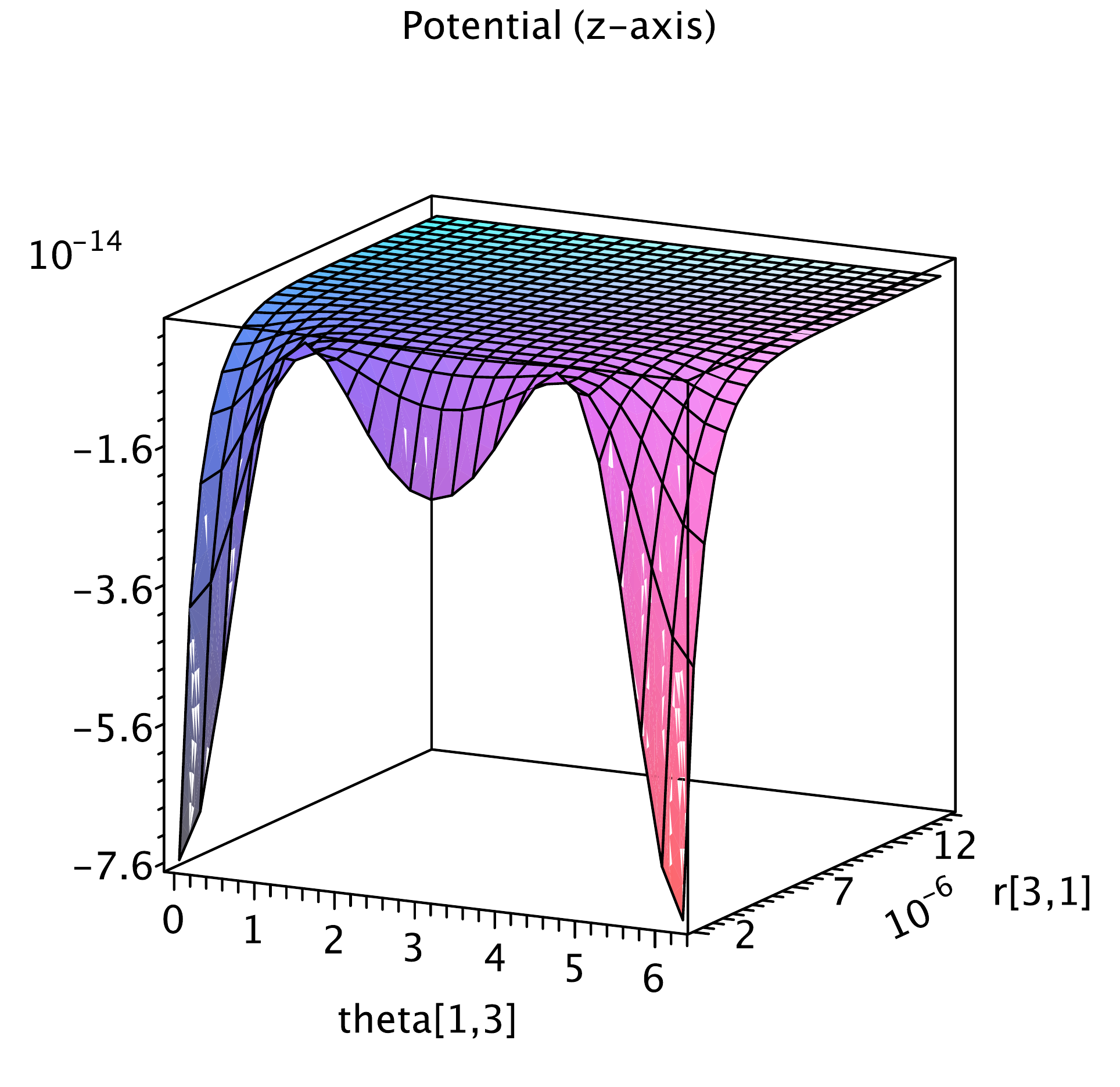}
\caption{A plot of the inter-sphere retarded potential (in units of $\hbar c/4\pi$) between three dielectric spheres in silicone fluid with $\epsilon_B=2.2$ at $T=293^{\circ}K$. Spheres 1 and 2 are held fixed along the z axis $10 R[1]$ apart. Here, the relative dielectric permittivity of the polystyrene spheres is $\epsilon_1=\epsilon_2=\epsilon_3=2.6$ . The angular momentum series representation has been truncated at $L_{truncate}=1$.}\label{fig:C3SFT}
\end{center}
\end{figure}

For simple setups we can evaluate Equation~(\ref{eq:NSPHEREFORCE2}) perturbatively. In the case of two spheres aligned along the z-axis one can calculate the force at $T=0$ as a multipole expansion and at $T>0$ by residues. An example plot is given in Figure~\ref{fig:TWOSPHERES}. In a similar fashion when the configuration consists of three spheres we can evaluate the three-body force (where the three two-body forces have been subtracted). In Figure~\ref{fig:C3SFT} a plot of the potential (derived from the integral of force with respect to the appropriate separation vectors) on one of the spheres as a function of one of the others position (the other being held fixed) is given.

\section{A Large $N$ Weak Coupling Limit}

Consider the case of $N$ identical spheres arranged in some fashion as $N\rightarrow \infty$, $\alpha^i =\alpha\rightarrow 0$, whilst $\lambda :=N\alpha_S$ is held fixed ($\alpha=\alpha_S \omega^3R^3/c^3$). Also assume that there is some representative separation, $s$, between the spheres and a corresponding orientation. In this case there are $(N-1)!$ irreducibly connected diagrams that contribute approximately equally to the $N$-body scattering. The expression for the associated potential reduces to (considering the case $T= 0$ to be specific)

\begin{equation}
\label{eq:LARGENSPHEREFORCE2}
V[1|N-1]\approx \pm (-1)^N\frac{\hbar c}{ N s}(N-1)!\int_{0}^{\infty}dX e^{-X}\left[  \alpha \bar{\mathcal{A}}^(X,s)\right]^N
\end{equation}
where $\mathcal{A}$ is a simple polynomial in $X$ obtained from the translation matrices $A$ after extracting the exponential prefactor. Taking the limit $N \rightarrow \infty$ we can extract the dominant $L=1$ term 
which reduces to (using Stirling's approximation)

\begin{equation}
\label{eq:LARGENSPHEREFORCE3}
V[1|N-1]\sim \pm (-1)^N\hbar c\frac{e^{-N}}{N^{3}}\lambda^N\frac{R^{3N}}{s^{1+3N}}.
\end{equation}
The use of this formula would be in adding an additional sphere to the system and measuring the resulting oscillation of the force compared to the absolute force before the sphere is added. Alternatively it could be used for quantifying the error of a truncated series representation for large numbers of weakly interacting particles.

\section{Conclusions}
\label{sec:conclusions}

In this talk I have summarised recent work we have done on calculating Casimir forces between spheres using a multiple scattering approach. This has been done at both zero and finite temperature. The total closed path length being always greater than the respective length scales of the associated radii is what enables the perturbative evaluation to be an accurate approximation. For large $N$ and weakly scattering spheres, we have deduced a scaling type formula for the force based on an average set of properties for the configuration. This could be useful for forces where we want to understand the behaviour as a function of $N$ rather than the detailed configuration. 

\ack
I would like to thank Stefan Buhmann, Stefan Scheel, Alex Crosse, Rachele Fermani and John Gracey for numerous helpful and constructive discussions. I would also like to thank the organisers of CEWQO09 for the opportunity to present this material. This work was supported by the SCALA programme of the European commission.

\section*{References}


\end{document}